\documentclass[12pt,a4paper]{article}

\usepackage{amsmath,amssymb,epsfig}
\begin{document}
\numberwithin{equation}{section}
\setlength{\unitlength}{.8mm}

\begin{titlepage} 
\vspace*{0.5cm}
\begin{center}
{\Large\bf Relativistic trajectory variables in $1 + 1$ 
dimensional Ruijsenaars-Schneider type models}
\end{center}
\vspace{2.5cm}
\begin{center}
{\large J\'anos Balog }
\end{center}
\bigskip
\begin{center}
{\it MTA Lend\"ulet Holographic QFT Group, Wigner Research Centre,\\
H-1525 Budapest 114, P.O.B. 49, Hungary}
\end{center}
\vspace{3.2cm}
\begin{abstract}
\noindent
A general algorithm to construct particle trajectories in $1+1$ dimensional
canonical relativistic models is presented. The method is a generalization
of the construction used in Ruijsenaars-Schneider models and provides a simple
proof of the fact that the latter satisfies the world-line conditions
granting proper physical Poincar\'e invariance. The 2-particle case for
the rational Ruijsenaars-Schneider model is worked out explicitly.
It is shown that the particle coordinates do not Poisson commute, as required
by the no-interaction theorem of Currie, Jordan and Sudarshan.
\end{abstract}

\end{titlepage}

\section{Introduction}

Relativistic particle physics remains almost completely 
synonymous with Relativistic Quantum Field Theory. Abandoning the particle 
alternative is due to an apparent conflict with relativistic causality in the
case of instantaneous action-at-a-distance interaction among point-like
particles. The famous ``no-interaction'' theorem of Currie, 
Jordan and Sudarshan \cite{1} states that in a canonical formalism if
we represent the 10 generators of the Poincar\'e group in terms of
canonical position and conjugate momentum variables, the Poincar\'e
Lie algebra relations exclude the presence of any non-trivial interaction.
This result is intuitively in line with the above mentioned verdict on 
incompatibility of causality with particle mechanics.

Soon after formulating the no-go theorem it was realized that the only 
assumption one has to give up is the canonical behaviour of the position 
variables and then a consistent fully Poincar\'e invariant theory describing 
the trajectories of an isolated system of point-like particles can be 
established. There are three, essentially equivalent formulations of the same 
theory. The first formulation is called Predictive Relativistic 
Mechanics~(PRM)~\cite{2} and is given by writing equations of motion in a 
Newtonian form
\begin{equation}
\ddot{x}^i_a=\mu^i_a(\{x\},\{\dot{x}\}),
\label{Newton}
\end{equation}
where $i = 1, 2, 3$ are space indices, $a = 1, 2,\dots,N$ are particle 
indices and the accelerations $\mu^i_a$ depend
on the instantaneous positions $x^i_a$ and velocities $\dot{x}^i_a$
of the particles. Relativistic invariance implies that the accelerations 
have to satisfy a set of quadratic, partial differential equations, 
the Currie-Hill (CH) equations \cite{3}. This ensures that if we transform
the particle trajectories (obtained by integrating the Newton equations 
(\ref{Newton})) into a Lorentz-boosted new coordinate system, the
particle accelerations in this new system are again satisfying the same
instantaneous Newton equations (as function of the positions and rapidities
in the new system). 

Unfortunately no explicit solution of the CH equations is known. There exist 
approximate solutions in the $1/c^2$ expansion ($c$ is the speed of light)
including the theory of classical electrons either in the 
Feynman-Wheeler~\cite{FW} or in Rohrlich's~\cite{Roh} formalism.
More important than this academic example are the equations of motion
describing compact binaries in General Relativity. These are known~\cite{CB1}
in the post-Newtonian (essentially $1/c^2$) expansion up to third order. The
equations are derived from General Relativity but some regularization
ambiguities related to the point-like approximation of the compact objects
are resolved \cite{CB2} by requiring that the CH equations are fulfilled
(approximately in the post-Newtonian expansion).

Given a solution of the CH equations a natural question is to construct
the 10 generators of the Poincar\'e group and ask if a symplectic structure
on the space of trajectories can be found such that the 10 quantities
generate the Poincar\'e group. It is not known if the 10 generators
always exist and if they are uniquely determined by the dynamics.

An alternative approach to relativistic mechanics \cite{5} 
can be called canonical. Here a phase space equipped with a symplectic 
structure is assumed from the beginning, together with the set of 
10 generators of the Poincar\'e group and the Hamiltonian of the model is 
identified with the generator of time translations from the Poincar\'e Lie 
algebra. In this approach the difficulty is to construct the particle 
positions $x^i_a$ (trajectory variables) as functions on the phase space. 
Given the initial positions the dynamics of the system determines the full 
space-time trajectories and the known action of the Poincar\'e group on the 
phase space tells us how the particle trajectories are transformed. In 
relativistic canonical mechanics we require that this induced action is 
identical to the usual linear Poincar\'e transformation of the space-time 
coordinates corresponding to the trajectories. For infinitesimal 
transformations the above consistency conditions require that the position
variables satisfy the Poisson-bracket relations
\begin{equation}
\{ P_i,x^j_a\}=-\delta^{ij},\qquad
\{ J^i,x^j_a\}=\epsilon^{ijk}x^k_a,\qquad
\{ K_i,x^j_a\}=-\frac{1}{c^2}x^i_a\dot{x}^j_a
\label{WLC}
\end{equation}
called the world-line conditions (WLC). Here $P_i$, $J^i$, $K_i$, 
respectively are the momentum, angular momentum, and Lorentz boost generators,
respectively, of the Poincar\'e group. If such particle coordinates are found,
their Poisson brackets
\begin{equation}
\{ x^i_a,x^j_b\}
\label{CR}
\end{equation}
must not vanish, otherwise, due to the no-go theorem, there is no interaction. 
Again, no explicit solution of the non-linear WLC equations is known. Most
constructions~\cite{5} are based on constraint dynamics and the trajectory 
variables are given only implicitly. Nevertheless the canonical approach 
has the advantage that only the trajectory variables $x^i_a$ have to be 
constructed, because the 10 integrals of the 
Poincar\'e group are there by construction from the beginning. 
Assuming the set $\{x^i_a\}$, $\{\dot{x}^i_a\}$ 
are good coordinates on the phase space (at least locally),
the accelerations occurring in (\ref{Newton}) can be calculated 
and by construction they must satisfy the Currie-Hill equations. 
There is also a third, covariant approach~\cite{6} to relativistic
mechanics, which is not discussed here.

Because of the lack of explicit solutions in $3 + 1$ space-time dimensions 
it is useful to study toy models in $1 + 1$ dimensions. Not many solutions
are known even for the $1+1$ dimensional analog of the CH equations.
Although the most general 2-particle solution has been found in $1 + 1$ 
dimensions~\cite{4}, but it is given in a very implicit form. 
In~\cite{CH} a completely explicit solution of
the Currie-Hill equations in $1 + 1$ dimensional Minkowski space-time
was presented. This solution can be written in terms of elementary functions
and provides an example in which important questions of the relativistic 
action-at-a-distance approach (conserved quantities, canonical structure, 
etc.) can be studied transparently.

The most famous $1+1$ dimensional examples are the exactly solvable 
Ruijsenaars-Schneider (RS) models~\cite{RS,R23}, the relativistic 
generalizations of the Calogero-Moser systems. 
The RS approach is canonical, and these
systems are important, because they are not only relativistic, but also 
integrable for any $N$ both classically and quantum-mechanically.
The original motivation for constructing the RS models 
was their relativistic invariance but later
the RS literature was almost entirely concerned with their 
integrable aspects. They have many applications in various fields
(see the review papers~\cite{reviews}). Trajectory variables 
satisfying (the $1+1$ dimensional analog of) the world-line 
conditions (\ref{WLC}) have been constructed 
but it is not clear if they are good coordinates on the entire phase
space and their explicit form in terms of the canonical variables
and their commutation relations (\ref{CR}) are not known.

Although the subject of RS models and their generalizations 
has a vast literature~\cite{reviews}, there are 
some open questions even in the case of the simplest RS models
(the question of physical non-relativistic limit, for instance). Even the
Poincar\'e invariance of the models has been questioned~\cite{BS}.
For this reason it is important to study these and related models
further.

In this paper we present a general algorithm to construct trajectory variables
satisfying the $1+1$ dimensional world-line conditions in canonical
relativistic models. The algorithm is a generalization of that used
in RS models~\cite{RS} but it is also useful for the RS models themselves. It
provides a simple proof of the fact that the WLC are satisfied
and this demonstrates true Poincar\'e invariance for this family of models.

Canonical relativistic mechanics in $1+1$ dimensional space-time
and the algorithm for the construction of trajectory variables are presented
in the next section. In section 3 we show how this general scheme is realized
in RS type models (from where the idea of the construction comes). In
section 4 the 2-particle case (for scattering type repulsive interaction)
is discussed. In this case the details can be worked out. In section 5
for the rational RS model, which is the simplest special case, the 2-particle
trajectories are explicitly calculated and in particular, the non-vanishing
of the Poisson bracket $\{x_1,x_2\}$ is shown. Finally there is a short 
summary in section~6.

\section{Canonical relativistic mechanics in $1+1$ dimensional
Minkowski space-time}

The starting point of canonical relativistic mechanics in $1+1$ dimensions
is a phase space equipped with a symplectic (Poisson) structure and the
set of 3 generators $\{{\cal H},{\cal P}, {\cal K}\}$ of the $1+1$ dimensional
Poincar\'e group satisfying
\begin{equation}
\{{\cal H},{\cal P}\}=0,\qquad\quad
\{{\cal H},{\cal K}\}={\cal P},\qquad\quad
\{{\cal P},{\cal K}\}=\frac{1}{c^2}{\cal H}.
\label{P11}
\end{equation}
We will associate differential operators $\hat A$ to functions $A$ on the
phase space in the usual way. Acting on any function ${\cal F}$ we have
\begin{equation}
\hat A{\cal F}=\{A,{\cal F}\}
\end{equation}
and in particular we will use the notation
\begin{equation}
\hat{\cal H}{\cal F}=\dot{\cal F},\qquad\qquad
\hat{\cal P}{\cal F}={\cal F}^\prime.
\end{equation}
The commutator of two such operators satisfies
\begin{equation}
[\hat A,\hat B]=\widehat{\{A,B\}}.
\end{equation}
For later use we note that a consequence of the Poincar\'e commutation
relations is the operator identity
\begin{equation}
\left(\hat{\cal K}+\frac{x}{c^2}\hat{\cal H}\right){\rm e}^{x\hat{\cal P}}
={\rm e}^{x\hat{\cal P}}\hat{\cal K}\,.
\label{ide}
\end{equation}
We will assume that the canonical coordinates $q_a$, $\theta_a$ satisfying
\begin{equation}
\{q_a,\theta_b\}=\delta_{ab},\qquad\quad a,b=1,2,\dots,N
\end{equation}
are good coordinates on our phase space and phase space functions will be
given as ${\cal F}(q,\theta)$.

The dynamics on the phase space is given by the Hamiltonian ${\cal H}$ and
we introduce the solution of the equations of motion $Q_a(t;q,\theta)$,
$T_b(t;q,\theta)$ satisfying
\begin{equation}
\frac{\partial}{\partial t}Q_a(t;q,\theta)=\dot q_a(Q,T),\qquad\quad
\frac{\partial}{\partial t}T_b(t;q,\theta)=\dot\theta_b(Q,T)
\end{equation}
and the initial conditions
\begin{equation}
Q_a(0;q,\theta)=q_a,\qquad\qquad T_b(0;q,\theta)=\theta_b.
\end{equation}
The time evolution of any function ${\cal F}$ is now solved by
\begin{equation}
\left({\rm e}^{t\hat{\cal H}}{\cal F}\right)(q,\theta)=
{\cal F}(Q,T).
\end{equation}
Quite analogously we introduce the space \lq\lq evolution'' generated
by the momentum~${\cal P}$
\begin{equation}
\left({\rm e}^{x\hat{\cal P}}{\cal F}\right)(q,\theta)=
{\cal F}(\bar Q,\bar T),
\end{equation}
where the space evolution is the solution of 
\begin{equation}
\frac{\partial}{\partial x}\bar Q_a(x;q,\theta)=
q^\prime_a(\bar Q,\bar T),\qquad\quad
\frac{\partial}{\partial x}T_b(x;q,\theta)=\theta^\prime_b(\bar Q,\bar T)
\end{equation}
and the initial conditions
\begin{equation}
\bar Q_a(0;q,\theta)=q_a,\qquad\qquad \bar T_b(0;q,\theta)=\theta_b.
\end{equation}

To specify particle dynamics we have to find the coordinates 
(trajectory variables) $x_a(q,\theta)$ for each particle. Their physical
meaning is the position of the $a^{\rm th}$ particle at $t=0$ and the
full trajectory is given by the evolution
\begin{equation}
x_a(t;q,\theta)=x_a(Q,T).
\end{equation}
We can calculate the velocity and acceleration of the particles:
\begin{equation}
v_a(q,\theta)=\dot x_a(q,\theta),\qquad\qquad
\mu_a(q,\theta)=\dot v_a(q,\theta)=\ddot x_a(q,\theta).
\end{equation}

The proper transformation property of the trajectory variables is obtained
from the requirement that by applying a Poincar\'e transformation generated
on the phase space by the generators in (\ref{P11}) the full space-time 
trajectories $x_a(t)$ have to transform by the standard linear Lorentz 
transformation formulas. These are called the world-line conditions   
and in the $1+1$ dimensional case are 
\begin{equation}
x^\prime_a=-1,\qquad\qquad \hat{\cal K}x_a=-\frac{1}{c^2}x_av_a,\qquad\quad
a=1,2,\dots,N.
\label{WLC11}
\end{equation}
(No summation over the particle index $a$ is implied.)

To construct a solution of (\ref{WLC11}) we have to associate to each
particle a Lorentz-invariant (boost-invariant) quantity $\rho_a$:
\begin{equation}
\rho_a:\qquad\quad \hat{\cal K}\rho_a=0
\label{bi}
\end{equation}
and find its space evolution
\begin{equation}
R_a(x)={\rm e}^{x\hat{\cal P}}\rho_a,\qquad\quad R_a(x;q,\theta)=
\rho_a(\bar Q,\bar T).
\end{equation}
Now the trajectory is defined as the solution of
\begin{equation}
R_a(x_a)=0.
\label{tr}
\end{equation}
This construction works if the solution of (\ref{tr}) exists and is unique.
If this is the case, we can take the derivative of it with respect to
any differential operator $\hat{\cal L}$:
\begin{equation}
\left(\hat{\cal L}R_a\right)(x_a)+R^\prime_a(x_a)\hat{\cal L}x_a=0.
\end{equation}
Choosing ${\cal L}={\cal P}$ we immediately get from here
\begin{equation}
\hat{\cal P}x_a=-1,
\end{equation}
i.e. the first world-line condition.
Further we get
\begin{equation}
\begin{split}
{\cal L}={\cal H}&\qquad\qquad \dot R_a+R^\prime_a\,\dot x_a=0,\\
{\cal L}={\cal K}&\qquad\qquad \hat{\cal K} R_a+R^\prime_a\,\hat{\cal K} x_a=0,
\end{split}
\end{equation}
where the argument of $R_a$ in the above formulas is $x=x_a$. If we now apply
(\ref{ide}) to $\rho_a$ and take it also at $x=x_a$ we see that the right 
hand side vanishes and we find
\begin{equation}
\hat{\cal K} R_a+\frac{x_a}{c^2}\dot R_a=0.
\end{equation}
Combining the last three equalities we get
\begin{equation}
R^\prime_a\,\hat{\cal K}x_a=\frac{x_a}{c^2}\dot R_a=-\frac{x_a}{c^2}
R^\prime_a\dot x_a
\end{equation}
and simplifying with the factor $R^\prime_a$ finally gives
the second world-line condition
\begin{equation}
\hat{\cal K}x_a=-\frac{x_a}{c^2}\dot x_a.
\end{equation}

\section{The Ruijsenaars-Schneider Ansatz}

Ruijsenaars and Schneider found a clever Ansatz \cite{RS,R23} for satisfying
(\ref{P11}):
\begin{equation}
{\cal H}=mc^2\sum_a\cosh\theta_a\,V_a,\qquad
{\cal P}=mc\sum_a\sinh\theta_a\,V_a,\qquad
{\cal K}=-\frac{1}{c}\sum_aq_a,
\label{Ansatz}
\end{equation}
where $m$ is the mass of the particles and
\begin{equation}
V_a=\prod_{b\not=a}f(q_a-q_b)
\end{equation}
is parametrized in terms of a positive, even function of one variable, $f(q)$.
From the relations in (\ref{P11}) the only nontrivial one is
\begin{equation}
\{{\cal H},{\cal P}\}=\frac{m^2c^3}{2}\sum_a\frac{\partial}{\partial q_a}
\prod_{b\not=a}f^2(q_a-q_b)=0.
\label{frel}
\end{equation}
For the two-particle case ($N=2$) this gives no further restrictions but for
$N>2$ the functional relations (\ref{frel}) are nontrivial. They are satisfied
if
\begin{equation}
f^2(q)=a+b\,{\mathfrak p}(q),
\label{Weier}
\end{equation}
where $a$ and $b$ are constants and ${\mathfrak p}(q)$ is the doubly periodic
Weierstrass function. Here we will study the degenerate cases where one
of the periods (type II, III) or both of them (type I) are sent to infinity 
and $f$ is characterized
by the positive, even \lq\lq pair potential'' $W$ as
\begin{equation}
f(x)=\sqrt{1+W(x)}.
\end{equation}
In the three degenerate cases we have
\begin{equation}
W(x)=\left\{
\begin{split}
\frac{g^2}{x^2}\qquad\quad\quad &{\rm type\ I\ (rational)},\\
\frac{\gamma^2}{\sinh^2\omega x}\qquad\quad &{\rm type\ II\ (hyperbolic)},\\
\frac{\gamma^2}{\sin^2\omega x}\qquad\quad &{\rm type\ III\ (trigonometric)}.
\end{split}
\right.
\end{equation}
In this paper we mainly focus on the cases I and II. Physically these cases
describe scattering with repulsive interaction and since the order
of particles cannot be changed the phase space is restricted to
\begin{equation}
q_1>q_2>\dots>q_N.
\end{equation}

Although the solution (\ref{Weier}) arose from the requirement of Poincar\'e
invariance, it turned out \cite{RS,R23} that the models (\ref{Ansatz}) are 
also Liouville integrable. This means that (beyond ${\cal H}$ and
${\cal P}$) there are further globally defined, commuting, conserved
quantities. Moreover, the corresponding action-angle variables can be found 
algebraically and the solution of the equations of motion can be given
explicitly.   

Next we discuss the nonrelativistic (nr) limit of the problem. For this purpose
we rescale the variables as
\begin{equation}
\theta_a=\frac{p_a}{mc},\qquad\qquad q_a=mcy_a.
\end{equation}
The nr variables are also canonical satisfying $\{y_a,p_b\}=\delta_{ab}$.
We also rescale the constant parameters as
\begin{equation}
\begin{split}
\omega=\frac{\mu}{mc},\qquad\qquad\gamma&=\sin\left(\frac{\mu g}{mc}\right)
\qquad{\rm \ \ (type\ II)},\\
\omega=\frac{\mu}{mc},\qquad\qquad\gamma&=\sinh\left(\frac{\mu g}{mc}\right)
\qquad{\rm (type\ III)}.
\end{split}
\end{equation}
Now we take the nr limit $c\to\infty$ and find
\begin{equation}
\lim_{c\to\infty}{\cal P}={\cal P}_{\rm nr}=\sum_ap_a,\qquad\qquad
{\cal K}=-m\sum_ay_a
\end{equation}
and
\begin{equation}
\lim_{c\to\infty}({\cal H}-Nmc^2)={\cal H}_{\rm nr}=
\frac{1}{2m}\sum_ap_a^2+\sum_{a<b}V(y_a-y_b), 
\end{equation}
where the nr potential is
\begin{equation}
V(x)=\left\{
\begin{split}
\frac{g^2}{mx^2}\qquad\quad\quad &{\rm type\ I},\\
\frac{\mu^2 g^2}{m\sinh^2\mu x}\qquad\quad &{\rm type\ II},\\
\frac{\mu^2 g^2}{m\sin^2\mu x}\qquad\quad &{\rm type\ III}.
\end{split}
\right.
\end{equation}
We see that the nr limit depends on the choice of parametrization (of both the
canonical variables and the constant parameters) and
it is not obvious if this formal $c\to\infty$ limit is what one could call
the physical nr limit (case of slowly moving particles). It was also 
questioned \cite{BS} if the RS models (which are also called 
relativistic Calogero-Moser type models) are truly describing relativistic
motion of interacting particles. Although the original motivation for
studying these models was their relativistic invariance, later mainly their
integrability aspects were in the focus of research and not the questions 
related to Poincar\'e invariant mechanics. 

The above mentioned doubts about true Poincar\'e invariance can be dispelled
by constructing the particle trajectories and showing that the world-line
conditions are satisfied. For RS models
the choice of the relativistic trajectory variables was motivated by
the fact that in a special case, for type II models with $\gamma=1$,
the model can be identified with the Sine-Gordon model and the particles
with Sine-Gordon solitons. This special case motivated the choice~\cite{RS}
\begin{equation}
\rho_a=q_a,\qquad\qquad R_a=\bar Q_a.
\end{equation}
It is obvious that this $\rho_a$ is boost invariant. Moreover,
\begin{equation}
\frac{\partial}{\partial x}\bar Q_a=q_a^\prime(\bar Q,\bar T)=
-mc\cosh\bar T_a\,V_a(\bar Q)<0,
\end{equation}
hence it is a monotonic function of $x$ and the solution of (\ref{tr})
is unique~\cite{RS}. Also
\begin{equation}
v_a=\dot x_a=\frac{\dot {\bar Q}_a}{\bar Q_a^\prime}=-c\tanh\bar T_a,
\end{equation}
hence
\begin{equation}
|v_a|<c
\end{equation}
as it should.

\section{Two-particle RS-type interaction}

In this section we will study the construction of trajectories more
explicitly in the two-particle case. For $N=2$ we have
\begin{equation}
\begin{split}
{\cal H}&=mc^2(\cosh\theta_1+\cosh\theta_2)f(q_1-q_2),\\
{\cal P}&=mc(\sinh\theta_1+\sinh\theta_2)f(q_1-q_2),
\end{split}
\qquad\quad
{\cal K}=-\frac{1}{c}(q_1+q_2).
\end{equation}
In the two-particle case it is useful to introduce the \lq\lq external''
variables
\begin{equation}
\zeta=q_1+q_2,\qquad\qquad \tau=\frac{\theta_1+\theta_2}{2}
\end{equation}
and the \lq\lq internal''  ones,
\begin{equation}
q=q_1-q_2,\qquad\qquad u=\frac{\theta_1-\theta_2}{2}.
\end{equation}
In terms of these,
\begin{equation}
{\cal H}=2mc^2\varepsilon\cosh\tau,\qquad
{\cal P}=2mc\,\varepsilon\sinh\tau,\qquad
{\cal K}=-\frac{1}{c}\zeta.
\end{equation}
Here
\begin{equation}
\varepsilon=\cosh uf(q)>1
\end{equation}
is the effective mass and $\tau$ is (up to a sign) the center of mass (COM)
rapidity. It is easy to see that $\varepsilon$ is Poincar\'e invariant:
\begin{equation}
\dot\varepsilon=\varepsilon^\prime=\hat{\cal K}\varepsilon=0.
\end{equation}
We also introduce the variable $\psi>0$ by
\begin{equation}
\varepsilon=\cosh\psi.
\end{equation}

We now want to study the dynamics of the system. First of all we see that
\begin{equation}
\dot\tau=0,\qquad\qquad \dot\zeta=-2mc^2\varepsilon\sinh\tau,
\end{equation}
i.e. $\tau$ is constant and $\zeta$ has linear time dependence. For the
internal variables $q$, $u$ we introduce the solution of the equations of
motion, $Q(t)$ and $U(t)$. To give them explicitly, we need
the following definitions. First of all, we assume that $W(q)$ is positive
and monotonically decreasing from $W(0)=\infty$ to $W(\infty)=0$. This is
satisfied in the type I and type II RS model, but since for $N=2$ there
is no restriction on the functional form of $W$, our considerations here 
are valid for any such function. Let us now define the function 
$g_\varepsilon(q)$ for $q>q_\varepsilon$ by
\begin{equation}
g_\varepsilon(q)=\int_{q_\varepsilon}^q\frac{{\rm d}y}{\sqrt{\varepsilon^2
-f^2(y)}},\qquad\quad f(q_\varepsilon)=\varepsilon.
\end{equation}
This is a monotonically increasing function and so is its functional
inverse $G_\varepsilon$:
\begin{equation}
G_\varepsilon(g_\varepsilon(q))=q.
\end{equation}
$G_\varepsilon(\xi)$ is monotonically increasing from $q_\varepsilon$ to
$\infty$ as $\xi$ goes from $0$ to $\infty$. We extend the domain
of definition of $G_\varepsilon(\xi)$ to $-\infty<\xi<\infty$ by requiring
it to be even. We will need the large $\xi$ asymptotics of 
$G_\varepsilon(\xi)$, which is of the form
\begin{equation}
G_\varepsilon(\xi)\approx \sqrt{\varepsilon^2-1}\xi+D(\varepsilon).
\end{equation}
The constant term $D(\varepsilon)$ will be used to characterize the 
time delay in the scattering process. For the type I (rational) RS model
\begin{equation}
G_\varepsilon(\xi)=\sqrt{(\varepsilon^2-1)\xi^2+\frac{g^2}{\varepsilon^2-1}}
\end{equation}
and in this case $D(\varepsilon)=0$. For the type II (hyperbolic) RS model
\begin{equation}
G_\varepsilon(\xi)=\frac{1}{\omega}{\rm arccosh} \left[
\cosh \omega q_\varepsilon \cosh(\omega\sqrt{\varepsilon^2-1}\xi)\right]
\end{equation}
and \cite{RS}
\begin{equation}
D(\varepsilon)=\frac{1}{\omega}\ln\cosh\omega q_\varepsilon=\frac{1}{2\omega}
\ln\left(1+\frac{\gamma^2}{\varepsilon^2-1}\right).
\end{equation}

The solution of the equations of motion is given by
\begin{equation}
Q(t)=G_\varepsilon(2mc^2\cosh\tau t-w),\qquad
U(t)=-{\rm arcsinh}\left(\frac{\dot Q}{2mc^2f(Q)\cosh\tau}\right),
\end{equation}
where
\begin{equation}
w={\rm sign}(u)g_\varepsilon(q).
\end{equation}

The solution of the ${\cal P}$ equations of motion is quite similar.
\begin{equation}
\tau^\prime=0,\qquad\quad\zeta^\prime=-2mc\,\varepsilon\cosh\tau,
\end{equation}
i.e. $\tau$ is constant and $\zeta$ has linear $x$ dependence. Furthermore,
defining the $x$-dependent $q$, $u$ as $\bar Q(x)$ and $\bar U(x)$, we find
\begin{equation}
\bar Q(x)=G_\varepsilon(2mc\sinh\tau x-w),\qquad
\bar U(x)=-{\rm arcsinh}\left(\frac{\bar Q^\prime}
{2mc f(\bar Q)\sinh\tau}\right).
\end{equation}
The $x$ evolution of the variables $q_1$ and $q_2$ is thus
\begin{equation}
\begin{split}
2\bar Q_1(x)&=\zeta-2mc\,\varepsilon\cosh\tau x
+G_\varepsilon(2mc\sinh\tau x-w),\\
2\bar Q_2(x)&=\zeta-2mc\,\varepsilon\cosh\tau x
-G_\varepsilon(2mc\sinh\tau x-w).
\end{split}
\end{equation}
Both solutions are strictly monotonically decreasing (from $+\infty$ to
$-\infty$) as $x$ goes from $-\infty$ to $+\infty$ and thus the equations
\begin{equation}
\begin{split}
\zeta-2mc\,\varepsilon\cosh\tau x_1
+G_\varepsilon(2mc\sinh\tau x_1-w)&=0,\\
\zeta-2mc\,\varepsilon\cosh\tau x_2
-G_\varepsilon(2mc\sinh\tau x_2-w)&=0
\end{split}
\label{x12}
\end{equation}
have unique solution for the trajectory variables $x_1$, $x_2$. It is
easy to see that
\begin{equation}
x_1>x_2.
\end{equation}
Since
\begin{equation}
\dot\zeta=-2mc^2\varepsilon\sinh\tau \qquad\quad{\rm and}\qquad\quad
\dot w=-2mc^2 \cosh\tau,
\end{equation}
the time evolution of the trajectory variables satisfy
\begin{equation}
\begin{split}
\zeta-2mc^2\varepsilon\sinh\tau t
-2mc\,\varepsilon\cosh\tau x_1(t)
+G_\varepsilon(2mc\sinh\tau x_1(t)+2mc^2\cosh\tau t-w)&=0,\\
\zeta-2mc^2\varepsilon\sinh\tau t
-2mc\,\varepsilon\cosh\tau x_2(t)
-G_\varepsilon(2mc\sinh\tau x_2(t)+2mc^2\cosh\tau t-w)&=0.
\end{split}
\end{equation}
This can be used to calculate the large $|t|$ asymptotics of the trajectories:
\begin{equation}
\begin{split}
x_1(t)\approx x_1^{(-)}(t)&=\frac{\tilde x_2}{\cosh\beta_1}+ct\tanh\beta_1+
\frac{\delta}{\cosh\beta_1},\\
x_2(t)\approx x_2^{(-)}(t)&=\frac{\tilde x_1}{\cosh\beta_2}+ct\tanh\beta_2-
\frac{\delta}{\cosh\beta_2},\\
\end{split}
\qquad (t\to -\infty)
\end{equation}
\begin{equation}
\begin{split}
x_1(t)\approx x_1^{(+)}(t)&=\frac{\tilde x_1}{\cosh\beta_2}+ct\tanh\beta_2+
\frac{\delta}{\cosh\beta_2},\\
x_2(t)\approx x_2^{(+)}(t)&=\frac{\tilde x_2}{\cosh\beta_1}+ct\tanh\beta_1-
\frac{\delta}{\cosh\beta_1},\\
\end{split}
\qquad (t\to +\infty)
\end{equation}
where
\begin{equation}
\tilde x_1=\frac{\zeta-\bar w}{2mc},\qquad
\tilde x_2=\frac{\zeta+\bar w}{2mc},\qquad
\bar w=\sqrt{\varepsilon^2-1}w,
\end{equation}
and
\begin{equation}
\beta_1=-(\psi+\tau),\qquad \beta_2=\psi-\tau,\qquad 
\delta=\frac{D(\varepsilon)}{2mc}.
\end{equation}
Classical scattering is characterized by the time delay defined by
\begin{equation}
x_2^{(+)}(t+\Delta t_1)=x_1^{(-)}(t),\qquad\qquad
x_1^{(+)}(t+\Delta t_2)=x_2^{(-)}(t)
\end{equation}
and is given by
\begin{equation}
c\Delta t_1=\frac{2\delta}{\sinh\beta_1},\qquad\qquad
c\Delta t_2=-\frac{2\delta}{\sinh\beta_2}.\qquad\qquad
\end{equation}
If we go to the COM frame where
\begin{equation}
c\tanh\beta_1=-\bar v,\qquad\qquad c\tanh\beta_2=\bar v,
\end{equation}
we find
\begin{equation}
\bar v\Delta t_1=\bar v\Delta t_2=-\sqrt{1-\frac{\bar v^2}{c^2}}
\frac{D(\varepsilon)}{mc}.
\end{equation}
Using the asymptotic rapidities $\beta_1$ and $\beta_2$ we can express
the energy and momentum of the two-particle system as
\begin{equation}
E=mc^2(\cosh\beta_1+\cosh\beta_2)={\cal H},\qquad
P=mc(\sinh\beta_1+\sinh\beta_2)=-{\cal P}
\end{equation}
i.e. in our conventions the physical momentum is $P=-{\cal P}$.

\section{Rational RS model}

It is not easy to find the solution of (\ref{x12}) in general. In this section
we consider the simplest nontrivial case, the type I (rational) RS model.
In this case $x_1$ and $x_2$ are the two solutions of the quadratic equation
\begin{equation} 
(\zeta-2mc\,\varepsilon\cosh\tau x)^2=(\varepsilon^2-1)(2mc\sinh\tau x-w)^2+
\frac{g^2}{\varepsilon^2-1}
\end{equation}
given by
\begin{equation}
x_1=\frac{p+\sqrt{p^2+Ah}}{A},\qquad\qquad
x_2=\frac{p-\sqrt{p^2+Ah}}{A},
\label{tr12}
\end{equation}
where
\begin{equation}
\begin{split}
A&=4m^2c^2(\cosh^2\psi+\sinh^2\tau),\\
p&=2mc(\zeta\cosh\psi\cosh\tau-\bar w\sinh\psi\sinh\tau),\\
h&=\bar w^2-\zeta^2+\frac{g^2}{\sinh^2\psi}.
\end{split}
\end{equation}

We see that the physical quantities are expressed in terms of the external
canonical variables $\zeta$ and $\tau$ and the new internal variables
$\bar w$ and $\psi$. The latter also form a canonically conjugate pair and
the non-vanishing Poisson brackets are
\begin{equation}
\{\zeta,\tau\}=1,\qquad\qquad \{\bar w,\psi\}=1.
\end{equation}
The Poincar\'e generators are given in terms of these variables as
\begin{equation}
{\cal H}=2mc^2\cosh\psi\cosh\tau,\quad
{\cal P}=2mc\cosh\psi\sinh\tau,\quad
{\cal K}=-\frac{1}{c}\zeta.
\end{equation}
Written in terms of these variables, the generators take a general, 
dynamics-independent form and all dynamics is encoded in the trajectories
(\ref{tr12}). These are equivalent to the relations 
\begin{equation}
x_1+x_2=\frac{2p}{A},\qquad\qquad x_1x_2=-\frac{h}{A}.
\label{sumprod}
\end{equation}

We can now write the Poisson bracket relations
\begin{equation}
\left\{x_1x_2,\frac{x_1+x_2}{2}\right\}
=\left\{\frac{p}{A},\frac{h}{A}\right\}
=-\left(\frac{x_1-x_2}{2}\right)\{x_1,x_2\}
\end{equation}
and by evaluating the $\{p/A,h/A\}$ Poisson bracket we find
\begin{equation}
\{x_1,x_2\}=-\frac{g^2}{m^3c^3(x_1-x_2)}\,\frac{\sinh\tau\cosh\psi}{
(\cosh^2\psi+\sinh^2\tau)^3}.
\end{equation}
We see that this Poisson bracket does not vanish, otherwise it would be
in contradiction with the no-interaction theorem.

The right hand side of the $\{x_1,x_2\}$ Poisson bracket is an 
expression that contains also the canonical variables. 
It would be nicer to write it in terms of
the physical variables $x_1$, $x_2$ and their time derivatives $v_1$, $v_2$.
This raises the question if $x_1$, $x_2$, $v_1$, $v_2$ form \lq\lq good''
coordinates on the phase space or at least in some part of the phase space.
To answer this question we supplement the relations (\ref{sumprod}) with
the time derivatives
\begin{equation}
v_1+v_2=\frac{2\dot p}{A}=-\frac{c\sinh 2\tau}{\cosh^2\psi+\sinh^2\tau}
\label{sumder}
\end{equation}
and
\begin{equation}
x_1v_2+x_2v_1=-\frac{\dot h}{A}=\frac{1}{m}\frac{
\bar w\sinh\psi\cosh\tau-\zeta\cosh\psi\sinh\tau}{\cosh^2\psi+\sinh^2\tau}.
\label{prodder}
\end{equation}
We see from (\ref{sumder}) that
\begin{equation}
\mu_1+\mu_2=0,
\end{equation}
i.e. the COM moves with constant velocity. We introduce the notation
\begin{equation}
u_1=\frac{v_1}{c},\qquad u_2=\frac{v_2}{c},\qquad v=\frac{u_1+u_2}{2}.
\end{equation}
Using the first equation in (\ref{sumprod}) and (\ref{prodder}) we can
express $\zeta$ and $\bar w$ in terms of the physical variables and the
asymptotic rapidities ($\psi$, $\tau$):
\begin{equation}
\begin{split}
\zeta&=(\cosh^2\psi+\sinh^2\tau)\frac{mc}{\cosh\psi}
[\cosh\tau(x_1+x_2)+\sinh\tau(x_1u_2+x_2u_1)],\\
\bar w&=(\cosh^2\psi+\sinh^2\tau)\frac{mc}{\sinh\psi}
[\sinh\tau(x_1+x_2)+\cosh\tau(x_1u_2+x_2u_1)].
\end{split}
\label{zetabarw}
\end{equation}
The variable $\psi$ is determined from (\ref{sumder})
\begin{equation}
\cosh^2\psi=-\frac{\sinh 2\tau}{2v}-\sinh^2\tau
\label{ch2psi}
\end{equation}
and finally from the second relation in (\ref{sumprod}), substituting the
expressions for $\zeta$, $\bar w$ and $\cosh^2\psi$, after some algebra, we get
a quintic equation satisfied by the variable $\xi=\tanh\tau$:
\begin{equation}
\xi^2(1+u_1u_2)+v\xi(1+\xi^2)=\frac{\lambda^2v^2}{(x_1-x_2)^2}
(1+v\xi)(1-\xi^2)^2,
\end{equation}
where $\lambda=\frac{g}{mc}$.
The solution of the quintic is further restricted by the requirements
\begin{equation}
{\rm sign}(\tau)=-{\rm sign}(v),\qquad\qquad
|\xi|>|v|,
\label{reqs}
\end{equation}
coming from (\ref{ch2psi}). 
An interesting problem is to find the subspace spanned by the variables
$x_1-x_2$, $u_1$, $u_2$ such that the quintic has unique solution also 
satisfying (\ref{reqs}) there. In this subspace also the accelerations
$\mu_1=-\mu_2$ can be expressed in terms of the instantaneous positions
$x_1$, $x_2$ and velocities $v_1$, $v_2$. The accelerations obtained
this way must satisfy the Currie-Hill equations. Unfortunately the 
accelerations are very complicated even though the type I RS model for two 
particles appears to be the simplest of all cases. The problem is drastically
simplified if we go to the COM frame where
\begin{equation}
\tau=0,\qquad\qquad \zeta={\rm const.}
\end{equation}
The following considerations are again valid for any pair potential $W(x)$, 
not just the one corresponding to the type I RS model. In this frame we have
\begin{equation}
x_1=\frac{\zeta+q}{2mc\,\varepsilon},\qquad\qquad
x_2=\frac{\zeta-q}{2mc\,\varepsilon},
\end{equation}
and consequently
\begin{equation}
x_1+x_2={\rm const.},\qquad\qquad
2x=x_1-x_2=\frac{q}{mc\,\varepsilon}.
\end{equation}
The physical equation of motion for this $x$ is obtained in two steps.
In the first step we have to solve
\begin{equation}
1-\frac{\dot x^2}{c^2}=\frac{1+W(2mc\,\varepsilon x)}{\varepsilon^2}
\label{vareps}
\end{equation}
for $\varepsilon$ and then the COM equations of motion are reduced to
the Newtonian form
\begin{equation}
\ddot x=-\frac{mc^3}{\varepsilon}W^\prime(2mc\,\varepsilon x).
\end{equation}
Again, it is not easy to find the solution of (\ref{vareps}) in general, but
for the type I RS model it can be done and we find for the acceleration
\begin{equation}
\ddot x=\frac{c^2x}{\lambda^2}(R-1)^2,
\end{equation}
where
\begin{equation}
R^2=1+\frac{\lambda^2}{x^2}\left(1-\frac{\dot x^2}{c^2}\right).
\end{equation}
The solution of this equation of motion is given by
\begin{equation}
x(t)=\frac{1}{\sinh2\psi}\sqrt{\lambda^2+(\sigma t)^2},\qquad\qquad
\sigma=2c\sinh^2\psi.
\end{equation}
The time variable $t$ is chosen such that $x(t)$ is minimal at $t=0$.
We see that the solution describes a scattering 
process with repulsive forces. The two particles, starting from
infinity, gradually approach each other and after the turning point, 
where the particles stop and reach the
minimal relative distance, are receding from each other.
We have seen in section 4 that for this model $D(\varepsilon)=0$, which means
that the time delay vanishes in this ``billiard ball'' type scattering process.

\section{Summary}

In the canonical approach to relativistic mechanics the construction
of models describing the motion of interacting particles consists of 
two main steps.
Assuming that the phase space is already equipped with a symplectic structure 
in the first step we have to find a suitable set of Poincar\'e generators
whose Poisson brackets form the Lie algebra of the Poincar\'e
group. Using these generators an action of the Poincar\'e group on the phase 
space can be constructed. The Hamiltonian of the model is identified with
the generator of time translations from the Poincar\'e Lie algebra.
In the second step the particle positions (trajectory variables) 
have to be found as functions on the phase space. Using the given time
evolution and starting from the given positions, the complete space-time 
trajectories of the particles can be determined. The action of the
Poincar\'e group on the phase space induces an action on the particle
trajectories and we require that this induced action is identical to
the usual linear Poincar\'e transformation formulas in terms of the 
space-time coordinates corresponding to the trajectories. This requirement,
for infinitesimal transformations, leads to consistency relations called
the world-line conditions. These are nontrivial, nonlinear relations
which must be satisfied by the trajectory variables. It is not known how to
satisfy the world-line conditions in general $3+1$ dimensional problems.

In this paper we presented a general method for solving the world-line
conditions in $1+1$ dimensional problems. All one has to do is to associate
a suitable Lorentz-invariant function on the phase space to each particle
and the method provides an equation the solution of which (provided its 
solution exists and is unique) gives trajectory variables satisfying the
world-line conditions.

Restricting attention to the two-particle problem, we constructed these
equations and demonstrated the existence and uniqueness of their solution
for a family of models, including the type I (rational) and 
type II (hyperbolic) Ruijsenaars-Schneider models and generalizations.
Further restricting attention to the simplest case, the rational RS model,
the trajectory variables were calculated explicitly by solving a quadratic
algebraic equation. This provides us with explicit formulas for the
two trajectory variables in terms of the original canonical coordinates
of the phase space. It would be desirable to use the physical variables
(particle position variables and their time derivatives) as coordinates
on the phase space (like in Newtonian mechanics) but it is not clear 
(even in this simple example)
what is the domain of these physical variables in which they can be used
as coordinates on the phase space and it is even more difficult to find
an explicit description of this inverse transformation. 

The above mentioned difficulties, which are present already in the very
simple model studied in this paper, illustrate the complexity of
the construction of relativistic particle models with interaction.
We hope to be able to return to these problems in a future publication.

 \vspace{1cm}
{\tt Acknowledgements}

\noindent 
This investigation was supported by the Hungarian National Science Fund 
OTKA (under K83267).


\end{document}